\documentclass[%
reprint,
 amsmath,amssymb,
 aps,
 prd
]{revtex4-2}

\usepackage{graphicx}
\usepackage{dcolumn}
\usepackage{bm}
\usepackage[dvips]{color}

\usepackage[utf8]{inputenc}
\newcommand{\be}{\begin{equation}}
\newcommand{\ee}{\end{equation}}
\newcommand{\ba}{\begin{eqnarray}}
\newcommand{\ea}{\end{eqnarray}}

\newcommand\spin[2]{s^{(#1)}_{#2}}

\definecolor{red}{named}{Red}

\begin{document}
\title{Neuromorphic quantum computing}
\author{Christian Pehle}
\email{christian.pehle@kip.uni-heidelberg.de}
\affiliation{%
Kirchhoff-Institute for Physics \\
Heidelberg University \\
Im Neuenheimer Feld 227, D-69120 Heidelberg
}%
\author{Christof Wetterich}
\email{c.wetterich@thphys.uni-heidelberg.de}
\affiliation{
Institute for Theoretical Physics \\
Heidelberg University \\
Philosophenweg 16, D-69120 Heidelberg
}%


\begin{abstract}
We propose that neuromorphic computing can perform quantum operations. Spiking neurons
in the active or silent states are connected to the two states of Ising spins. A quantum
density matrix is constructed from the expectation values and correlations of the Ising spins.
As a step towards quantum computation we show for a two qubit system that quantum gates can be learned
as a change of parameters for neural network dynamics. Our proposal for
probabilistic computing goes beyond Markov chains, which are based on transition probabilities.
Constraints on classical probability distributions relate changes made in one part of
the system to other parts, similar to entangled quantum systems.
\end{abstract}

\maketitle

\section{Introduction}

It has been suggested that artificial neural networks or neuromorphic computers can perform the operations necessary for quantum computing
\cite{wetterich:2018, pehle:2018}. It has already been argued that several important tasks within quantum computation
can be performed by neural networks \cite{carleo:2016, torlai:2018, sharir:2020} and frameworks exist
which enable the implementation of several of these techniques (c.f. \cite{broughton2020tensorflow} and references therein). Beyond this, the proposal of
\cite{wetterich:2018, pehle:2018} suggests that a full quantum computation  may be possible
by use of stochastic information in a system of neurons. The present work proposes for 
a first time a concrete implementation of quantum operations by spiking neurons.

We demonstrate four important steps in that direction:
\begin{enumerate}
    \item We describe spiking neurons by a system of differential equations with parameters $W$. Expectation values and correlations of appropriate Ising spins can be extracted
    from the dynamics of the interacting neurons. They depend on the parameters $W$. We demonstrate that for every given two-qubit density matrix $\rho$ the system of neurons can
    adapt or learn suitable parameters $W$ such that the correlation map of expectation values and correlations of Ising
    spins expresses the density matrix.
    \item We show that the correlation map \cite{wetterich:2018} from expectation values and correlations of classical Ising spins
    to a quantum density matrix is complete in the case of two qubits.  This means that for every
    possible quantum density matrix $\rho$ there exists a probability distribution $p$ for the
    configurations of classical Ising spins, such that $\rho$ is realized by the correlation map. This
    completeness was not proven before.
    \item We establish how unitary transformations of $\rho$ corresponding to arbitrary quantum gates can be performed by a change of probability distributions $p$.
    \item We demonstrate that the minimal correlation map is not a complete bit-quantum map
    for three or more qubits. We propose an extended correlation map.
\end{enumerate}

Our approach centers on the use of expectation values and particular correlations of Ising spins for the construction
of quantum density matrices. Correlated Ising systems are efficient models for many biological systems \cite{hopfield:1982}. We concentrate first
here on two qubits and discuss the scaling to a higher number of qubits at the end.

\begin{figure}
\includegraphics[]{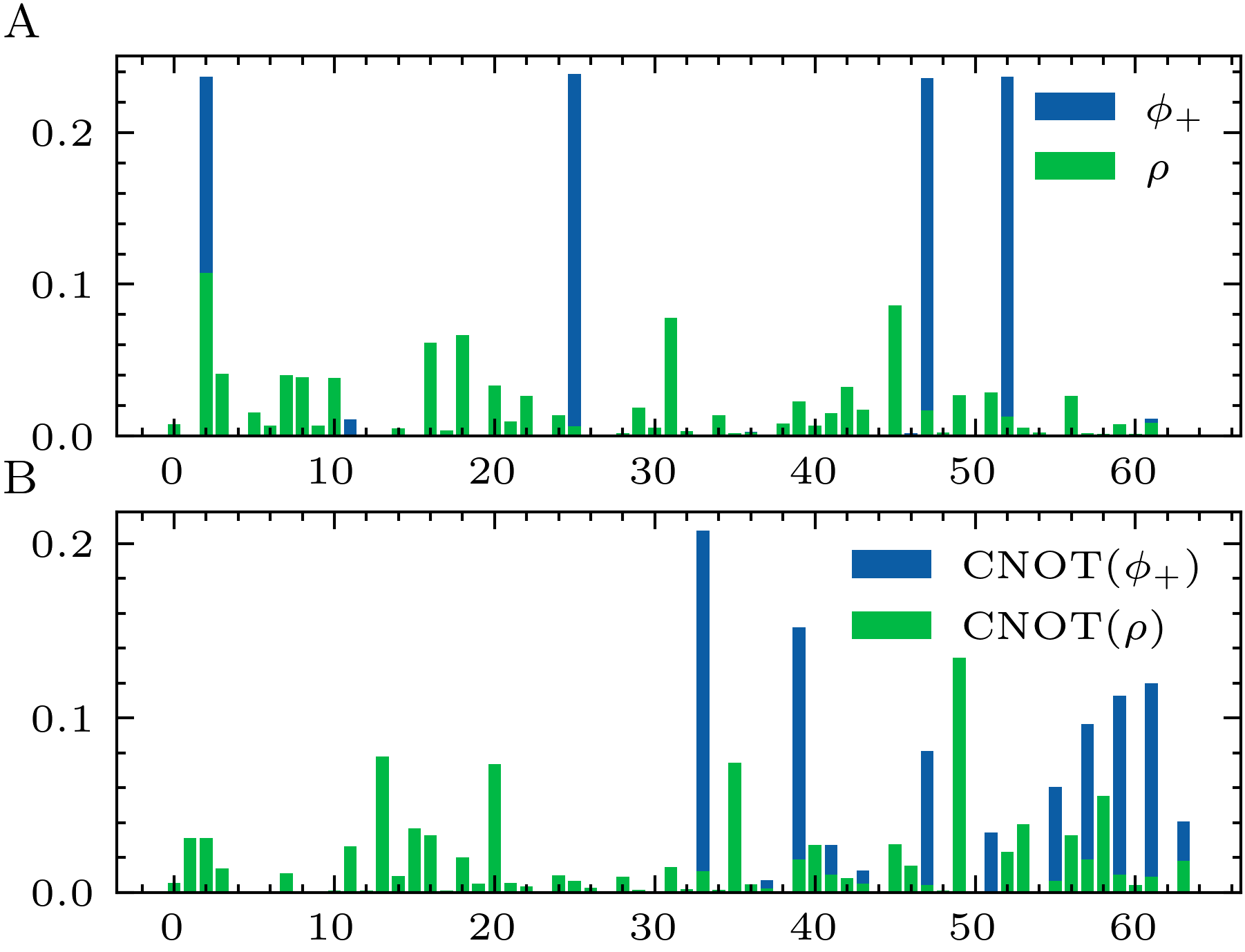}
\caption{\label{figure:probabilities} Classical probabilities for quantum states. We indicate the probabilities corresponding to the density
matrices of the maximally entangled pure state $\psi_+ = \frac{1}{\sqrt{2}} (\lvert 0 0 \rangle + \lvert 1 1 \rangle)$ (blue) and a randomly generated density matrix $\rho$ (A) (green).
We also show their transformation under a CNOT gate (B). The labels $0\ldots 63$ are best thought of as bit vectors and label the state of $6$ classical Ising spins $s^{(i)}_k, i=1,2, k=1\ldots 3 .$ For example the label $3$  corresponds to the spin state $[-1,-1,-1,-1,1,1]$. This figure
demonstrates that entangled quantum states can be realized by classical probability distributions, and quantum
gates by changes of these probability distributions.
}
\end{figure}

The quantum operations are performed by changes of the probability distributions
and corresponding changes of the expectation values and correlations. These changes
cannot be realized by Markov chains for which otherwise deterministic operations
are performed with certain "transition probabilities". Probabilistic computing
beyond Markov chains opens a rich area of new possibilities.

\section{Correlation map for two qubits}
The correlation map for a two qubit quantum system involves six classical Ising spins \cite{wetterich:2018} that are
grouped in two pairs of three Ising spins $s^{(i)}_k$. Here $i = 1,2$ corresponds to the two quantum spins,
and $k = 1 \ldots 3$ is a associated to the three cartesian directions of a given quantum spin. We can form a real $4 \times 4$ matrix $\chi$ of expectation values
and correlations as
\begin{align}\label{eq:exp_correlations}
\chi_{00} = 1 , \chi_{0k} = \langle s^{(1)}_k \rangle, \chi_{l0} &= \langle s^{(2)}_l \rangle, \chi_{kl} = \langle s^{(1)}_k s^{(2)}_l \rangle.
\end{align}
If we denote by $\tau_k, k = 1 \ldots 3$ the three Pauli matrices and by $\tau_0$ the identity matrix, we can define
the $U(4)$-generators
\begin{align}
L_{\mu \nu} &= \tau_\mu \otimes \tau_\nu, \quad \mu = 0\ldots  3, \quad \nu = 0 \ldots 3.
\end{align}
The bit quantum map organises the expectation values and correlations (\ref{eq:exp_correlations}) into
a density matrix
\begin{align}\label{eq:density_from_correlations}
\rho &= \frac{1}{4} \chi_{\mu \nu} L_{\mu \nu}.
\end{align}
(Summation over double indices is implied.) We denote this map by
\begin{align}
f \colon \mathbf{R}^{4 \times 4} \to \mathbf{C}^{4 \times 4},\quad \chi \mapsto  \rho = \frac{1}{4} \chi_{\mu \nu} L_{\mu \nu}.
\end{align}

\begin{figure}
\includegraphics[]{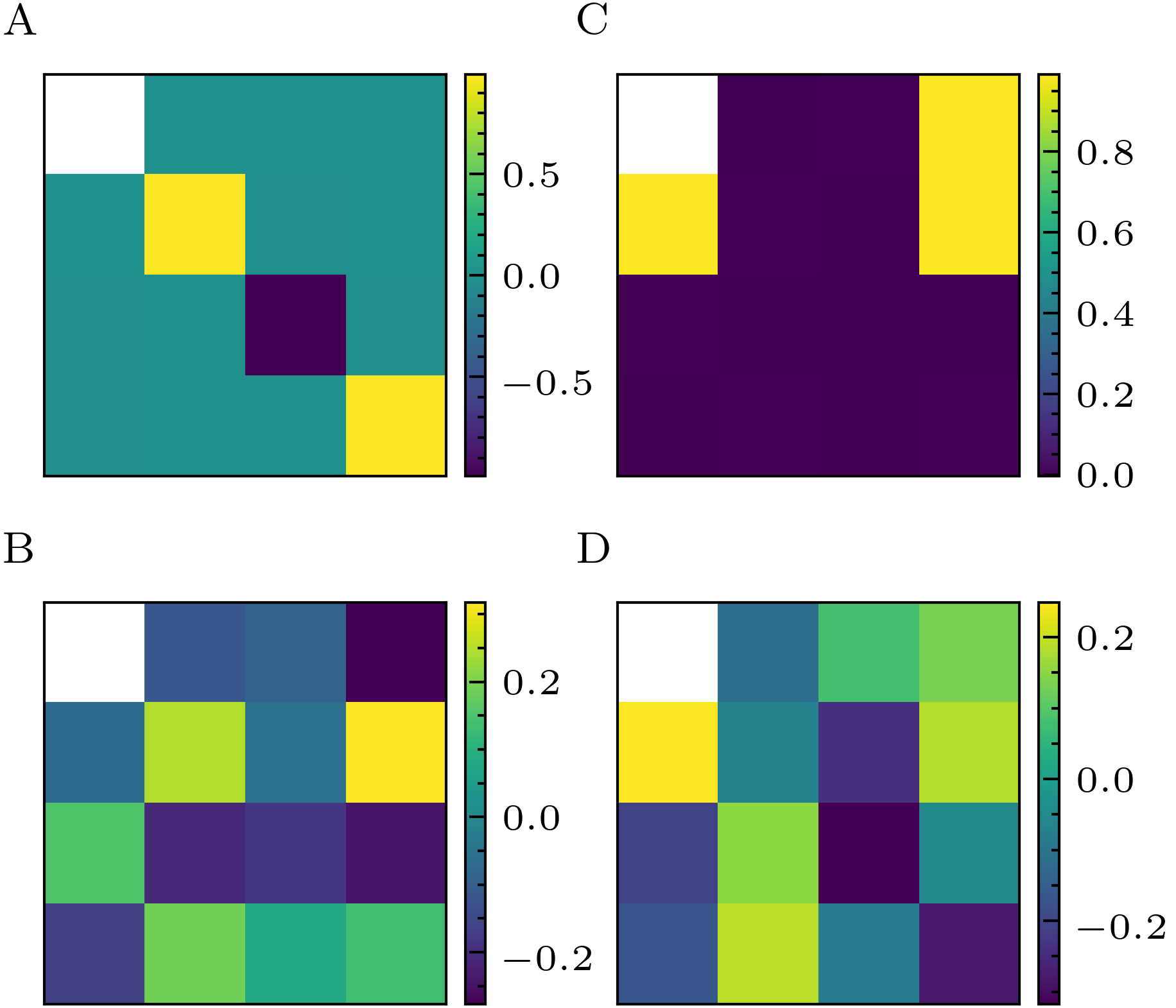}
\caption{We show the $ 15 = 2 \times 3 + 9$ spin
 expectation values $\langle s^{(1)}_k \rangle$ and 
 correlations $\langle s^{(1)}_k s^{(2)}_l \rangle$ 
 corresponding to $\psi_+$ (A) and $\rho$ (B).
The first row contains the three expectation values $\langle s^{(1)}_k \rangle$,
the first column the three expectation values $\langle s^{(2)}_k \rangle$. 
The remaining $3 \times 3$ entries are made up of the correlations $\langle s^{(1)}_k s^{(2)}_l \rangle$. 
The transformed spin expectation values and correlations related to the density matrices $\text{CNOT}(\psi_+)$ 
 and $\text{CNOT}(\rho)$  
are shown in (C) and (D) respectively. Note the different color scale between (A)(B) and (C)(D).}
\label{fig:task_1_expectations}
\end{figure}

The density matrix $\rho$ is a complex hermitian
$4 \times 4$ matrix with $\text{tr}(\rho) = 1$. A quantum density matrix has to be positive - all eigenvalues $\lambda$ of $\rho$ have to obey $\lambda \geq 0$.
This imposes restrictions on the classical probability distribution $p$ or the expectation values (\ref{eq:exp_correlations})
that can realize a quantum density matrix. These restrictions are called the "quantum constraints". 

The states $\tau = 0 \ldots 63$ of the classical spin system correspond to the $2^6 = 64$ configurations of
Ising spins $\{ s^{(1)}_1, s^{(1)}_2, s^{(1)}_3, s^{(2)}_1, s^{(2)}_2, s^{(2)}_3\}$ that can take the values $s = \pm 1$. The probabilistic
system is defined by associating to each $\tau$ a probability $p_\tau \geq 0, \sum_\tau p_\tau = 1.$ The expectation values (\ref{eq:exp_correlations}) are 
formed in the standard way, multiplying the values of $(s_\gamma)_\tau$ or $(s_\gamma)_\tau (s_\delta)_\tau$ in a given state $\tau$ with $p_\tau$ and summing over $\tau$. This
results in a weighted sum
\begin{align}\label{eq:classical_exp_and_correlations}
\chi_{k0} = \sigma^{(k 0)}_\tau p_\tau,\quad \chi_{0k} = \sigma^{(0k)}_\tau p_\tau,\quad \chi_{kl} = \sigma^{(kl)}_\tau p_\tau,
\end{align}
with signs $\sigma^{(a)}_\tau = \pm 1$ given by
\begin{align}
\sigma^{(k0)}_\tau &= (-1)^{1 + \text{bin}(\tau)[k]},\\ 
\sigma^{(0k)}_\tau &= (-1)^{1 + \text{bin}(\tau)[3 + k]},\\
\sigma^{(kl)}_\tau &= (-1)^{1 + \text{bin}(\tau)[k]} (-1)^{1 + \text{bin}(\tau)[3+l]}.
\end{align}
Here $\text{bin}(\tau)$ denotes the 6 bit binary representation of a number $0\ldots 63$ and
$\text{bin}(\tau)[k]$ an the entry at index $k$ of that vector. For the example $\tau = (1,-1,-1,1,1,1)$ one
has $\text{bin}[\tau] = (1,0,0,1,1,1)$, $\text{bin}(\tau)[2] = 0$ reads out the value at the second place
and $\sigma_\tau^{(20)} = -1$ is the value  of the spin $s^{(1)}_2$ in the state $\tau$. We denote this map by
\begin{align}
g \colon \mathbf{R}^{64} \to \mathbf{R}^{4 \times 4},\quad p \mapsto \chi.
\end{align}

The bit quantum map $b$ can be seen as a map from the classical probability distribution $p$ to
the quantum density matrix,
\begin{align}\label{eq:bit_quantum_map}
b = g \circ f \colon \mathbf{R}^{64} \to \mathbf{C}^{4 \times 4}, \quad p \mapsto \rho.
\end{align}
It has the property that the particular classical expectation values and correlations (\ref{eq:classical_exp_and_correlations}) coincide with
the expectation values of quantum spins in the cartesian directions, and the corresponding quantum correlations. The density matrices can be constructed
(or reconstructed) by use of the particular classical or quantum correlations. This is analogous to the reconstruction of the density matrix for photons by 
appropriate correlations \cite{james:2001,paris:2004,kiesel:2007}. Examples for probability distributions $p$ realizing particular quantum density matrices $\rho$ are shown in figure \ref{figure:probabilities}.
In fig. \ref{fig:task_1_expectations} we indicate the corresponding spin expectation values and correlations.

\section{Completeness of the correlation map for $Q=2$}
Our first task is a demonstration that the bit quantum map $b$ is complete, in the sense that for 
every positive hermitian normalized quantum density matrix there exists at least one 
classical probability distribution $p$ which realizes $\rho$
by use of equation (\ref{eq:bit_quantum_map}). For this task it will be 
convenient to work with the "classical wave function" $q$ which is a type of probability
amplitude. Its components $q_\tau$ are related to $p_\tau$ by
\begin{align}\label{eq:p_defintion}
p_\tau = \frac{q_\tau^2}{\sum_\tau q^2_\tau}.
\end{align}
This defines one further map
\begin{align}
h \colon \mathbf{R}^{64} \to \mathbf{R}^{64},\quad q \mapsto p = \frac{q^2}{\lvert q \rvert^2_2}.
\end{align}

Changes of the probability distribution correspond to rotations of the vector $q$, with $p_\tau > 0$ 
and $\sum_\tau p_\tau = 1$ guaranteed by the construction
(\ref{eq:p_defintion}). We need to find for each given density $4 \times 4$ 
matrix $\rho \in \mathbf{C}^{4 \times 4}$ a vector $q \in \mathbf{R}^{64}$, 
such that it maps to the density matrix $\rho$ under the composition $b \circ h$.
 Clearly any such $q$ is not unique. 

\begin{figure}
\includegraphics{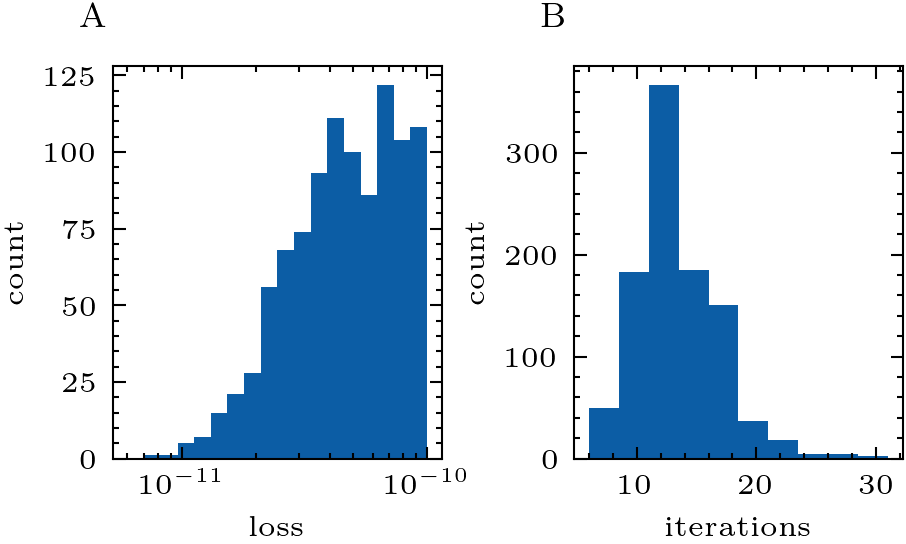}
\caption{\label{figure:2} Optimization results for finding a vectors $q_i \in \mathbf{R}^{64}$, such the loss $l_{\rho_i}(q_i)$ (\ref{eq:q_loss}) is minimized. We generate $10^3$ density matrices $\rho_i$ by starting from a randomly generated density matrix $\rho_0$ and iteratively apply three unitary transformations $\text{CNOT}$, Hadamard $H \otimes I$ and rotation $R_{1/8} \otimes I$. The optimization is stopped once the loss falls below $10^{-10}$. Figure (\textbf{A}) is a histogram of the resulting final losses, figure (\textbf{B}) is a histogram of the required iterations. 
}
\end{figure}

In order to find one such $q$ we minimize
\begin{align}\label{eq:q_loss}
l_\rho(q) = \lvert (b \circ h)(q) - \rho \rvert_2^2
\end{align}
by gradient descent on $q$. To verify numerically that this approach works we performed the following test: Starting with a randomly generated density matrix $\rho_0$, we iteratively apply the three unitary transformations $\text{CNOT}$, Hadamard $H \otimes I$ and rotation $R_{1/8} \otimes I$ to obtain density matrices $\rho_i$. The quantum gates preserve the positivity of $\rho$. Applying them many times leads to a dense covering to the space of quantum density matrices, which becomes a full covering if the number of steps goes to infinity. We further explore the space of $\rho$ by investigating different $\rho_0$. For each $\rho_i$ we solve the optimization problem with
loss $l_{\rho_i}(q)$ to obtain corresponding vectors $q_i$. The results of this optimization are shown in figure \ref{figure:2}. For all $\rho_i$ we find vectors $q$ realizing the bit quantum map $b \circ h$ with high precision already after rather few iterations. We conclude that for two qubits the bit quantum map is complete.

%
%

Considering one of the vectors $q$ found in this way as a representation of $\rho$ , 
our second step asks how a given unitary transformation of $\rho$ can be represented 
as a transformation of $q$. Considering $q$ as a classical probabilistic object, we ask how the
classical system can learn a quantum operation. The goal is to find for any vector $q \in \mathbf{R}^{64}$
 a matrix $M \in \mathbf{R}^{64 \times 64}$, such that the transformed vector $M q$ yields the density 
 matrix $U \rho U^\dagger$ related to $\rho$ by a given unitary transformation $U$. A 
simple "learning" or optimization goal consists in minimizing the Frobenius norm between 
the two density matrices $U \rho U^\dagger = U (b \circ h)(q) U^\dagger \equiv  U((b \circ h)(q))$ 
and $(b \circ h)(M q)$,
\begin{align}
l_U(M) = \rvert \text{U}((b \circ h)(q)) - (b \circ h)(M q)\lvert_2^2
\end{align}
Here we use the notation $U(\rho) = U \rho U^\dagger.$ 

This minimization can again be implemented using gradient descent. In figures \ref{figure:probabilities}, \ref{fig:task_1_expectations} we show the resulting probabilities \begin{align}
p = h(q),\quad p' = h(M q),
\end{align}
as well as the corresponding matrices of expectation values and correlations 
\begin{align}
\chi = (g \circ h)(q),\quad \chi' = (g \circ h)(M q),
\end{align}
when this minimization procedure is applied to two example initial density matrices and their 
associated $q$ vectors.
 The unitary transformation $U$ is taken to be $\text{CNOT}$. We have investigated 
 many different $\rho$ and always found a 
satisfactory $M$ with this procedure. Thus the classical system "learns" the matrix $M$ 
necessary for a unitary quantum gate. 
 We emphasize that $M$ depends on $q$, such that $Mq = M(q)q$ is a 
 non-linear map \cite{wetterich:2018}.

\section{Quantum computing with spiking neurons}
\begin{figure}
\includegraphics{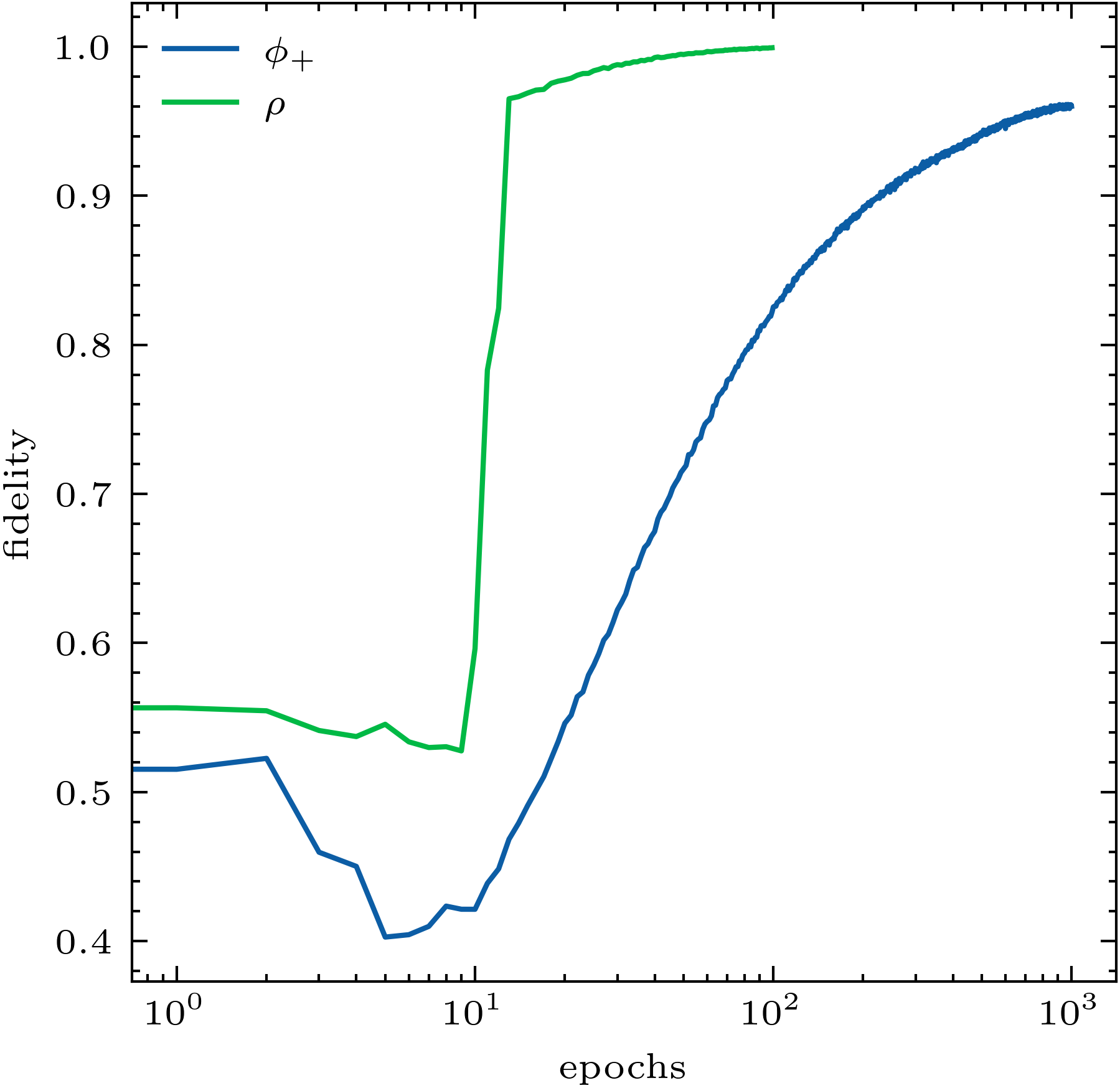}
\caption{Fidelity of density matrices obtained 
by correlations of spiking neurons as a function
of training epochs. Shown are approximation results 
for the density matrix of a maximally entangled pure state
 $\phi_+ = \frac{1}{\sqrt{2}} (\lvert 11\rangle + \lvert 00\rangle)$
and a random density matrix $\rho$. 
}
\label{figure:spike_loss}
\end{figure}

We finally turn to our third task of implementation by neuromorphic computing. So 
far the Ising spins $s$ entered only indirectly through their state probabilities $p$
and associated correlations and expectation values $\chi$. We next want to study classical systems for which correlations and expectation values can be determined from temporal averages. 
In the context of neuroscience there is a long history of applying spin-glass models to the study of biological neural networks \cite{hopfield:1982}.
 The general idea is to consider each biological neuron to have two states \emph{active} and \emph{silent}. A neuron is considered active (or refractory) if 
 it has produced a \emph{spike} within a certain sampling time window and silent otherwise. Within this framework, experimental work has been carried out to study pairwise and higher
  correlations of biological neurons (e.g. \cite{schneidman:2006, ohiorhenuan:2010}).

One immediate way of obtaining the state probabilities $p$ is from \emph{neural sampling} \cite{buesing:2011, petrovici:2016}.
Since we need only the expectation values and correlations (\ref{eq:exp_correlations}) we directly focus on these quantities and do not aim to 
resolve the probability distributions completely. The six spin variables $s^{(i)}_k(t)$ are associated to six particular neurons in a larger
network, with $s = 1$ if the neuron is active (refractory) and $s = -1$ if it is silent. Expectation values can be formed by measuring
the duration of active and silent states,
\begin{align}
\label{eq:spin_average_1}
\langle s^{(i)}_k \rangle &= \frac{1}{T} \int_0^T s^{(i)}_k(t) \text{dt}, \\ 
\label{eq:spin_average_2}
\langle s^{(1)}_k s^{(2)}_l \rangle &= \frac{1}{T} \int_0^T s^{(1)}_k(t) s^{(2)}_l(t) \text{dt}.
\end{align}
Instead of spin variables $s$ which take values $\{1, -1\}$, we want to consider variables $z$ with values in $\{1, 0\}$, $s = 2z - 1$. A given selected neuron has the value $z = 1$ during the refractory period (active state), and $z = 0$ otherwise (silent state). We choose a simple so called \emph{leaky-integrate and fire} (LIF) neuron model. The model is specified by $3 n$ state variables $v_i, I_i, r_i, i = 1\ldots n$, called the membrane voltages, synaptic currents and refractory states. The dynamics has phases of continuous evolution and jumps or spikes. The continuous evolution obeys
the differential equations
\begin{align}
\label{eq:voltages}
\dot{v} &= (1 - \Theta(r))(v_{\text{l}} - v + I) \\
\dot{I} &= -I + I_{\text{in}} \\
\label{eq:refractory}
\dot{r} &= -\frac{1}{t_\text{refrac}} \Theta(r).
\end{align}
Here $\Theta$ denotes the Heaviside function and multiplication in (\ref{eq:voltages}) is pointwise for every $i$ separately. 
Characteristic for a spiking neuron model are the jumps or discrete changes of the 
state variables $v,I,r$ at particular times. 
Whenever one of the membrane voltages $v_i$ reaches a
 threshold $(v_\text{th})_i$ during the continuous evolution
\begin{align}\label{eq:jump_condition}
v_i - (v_{\text{th}})_i = 0,
\end{align}
the state variables undergo discontinuous jumps.
The transition equations
\begin{align}\label{eq:transition_equations}
v^+ &= v^- + \xi (v_{r} - v_{\text{th}}), \\
\label{eq:transition_eq_current}
I^+ &= I^- + W_{\text{rec}} \cdot \xi,\\
\label{eq:transition_eq_refractory}
r^+ &= r^- + \xi,
\end{align}
specify the state before and after the transition, $v^\pm = v(t^\pm), I^\pm = I(t^\pm)$ and $r^\pm = r(t^\pm)$. 
Here $\xi \in \mathbf{R}^n$ is a binary vector, $\xi_i = 1$ for the value of $i$ for which the threshhold voltage 
is reached and (\ref{eq:jump_condition}) obeyed, and $\xi_j = 0$ for $v_j \neq (v_\text{th})_j$. In equation (\ref{eq:transition_equations})
the multiplication is pointwise such that only the voltage of the spiking neuron changes. 
In (\ref{eq:jump_condition}-\ref{eq:transition_eq_refractory}) $v_{\text{th}}, v_{r}, v_{l} \in \mathbf{R}^n$ 
are the \emph{threshold}, \emph{reset} and \emph{leak} potential.
Finally $I_\text{in}$ is an input current and $W_{\text{rec}}$ is a matrix in $\mathbf{R}^{n \times n}$ which
parameterises the response of the currents $I_j$ of all neurons to the "firing" of the spiking neuron $i$. 

Equation (\ref{eq:refractory}) specifies a linear decrease of $r_i$ until $r_i = 0$ is reached. As long as $r_i > 0$ the neuron is
considered to be active. The neuron remains active for a refractory period $t_\text{refrac}$ after its firing at $t_k$. For this period
its voltage does not change, 
\begin{align}
\dot{v}_i([t_k, t_k + t_{\text{refrac}}]) = 0,
\end{align}
as implied by equation (\ref{eq:voltages}).

Ising spins $s(t)$ or the associated occupation numbers $z(t)$ are defined by
\begin{align}
    z(t) = \Theta(r(t)),
\end{align}
which is one during the refractory period and zero otherwise.
Six selected neurons define the Ising spins $s^{(i)}_k = 2 z^{(i)}_k - 1$ for which the expectation values
and correlations (\ref{eq:spin_average_1}),(\ref{eq:spin_average_2}) define the quantum density matrix
$\rho$ by equations (\ref{eq:exp_correlations}), (\ref{eq:density_from_correlations}).

The last quantity to be specified is the input current to the
 $n$ LIF neurons. We model spike input from $m$ additional input spike sources.
 At times $t_l$ the source neuron $q_l$ fires. The response of the $n$ LIF neurons is
 given by
\begin{align}
(I_\text{in})_j = \sum_l (W_{\text{in}})_{j,q_l} \delta(t - t_l)\quad q_l \in  {1, \ldots, m}.
\end{align}
Here $\delta(t - t_l)$ is the Dirac delta distribution, such that we model an 
immediate response of all $n$ LIF-neurons to every input spike at $t_l$.
 The $n \times m$ matrix $W_\text{in} \in \mathbf{R}^{n \times m}$
parametrises the height of the jump in the currents $I_j$ upon arrival of a spike at input source $q_l$. In 
our experiments the arrival times $t_l$  of the input spikes are the result of $m$ independent Poisson 
point processes with rates $\lambda_q, q = 1\ldots m$. 

By solving the differential equations with 
jumps for given input spikes we can compute $z_j(t)$ and therefore
\begin{equation}
s^{(i)}_k(t) = 2 z^{(i)}_k(t) - 1.
\end{equation}
In the experiments we carried out we took $n > 6$. The six spin variables $s^{(i)}_k$ are 
recovered by projecting to the first six components
\begin{equation}
\pi \colon \mathbf{R}^n \to \mathbf{R}^6, (s_1, \ldots, s_n) \mapsto (s_1, \ldots, s_6).
\end{equation}

Given a density matrix $\rho$, we can now formulate an optimization problem for $W_{\text{in}}, W_{\text{rec}}$. One obstacle in doing so is that the jumps introduce discontinuities, which need to be taken into account. We choose here to solve this issue by introducing a smooth approximation to the heaviside function $\Theta_\epsilon$. More specifically we use a fast sigmoid as in \cite{zenke:2018} with parameter $1/\epsilon = 100$. This is a common way in which spiking neuron models can be made amendable to gradient descent optimization \cite{esser:2016, zenke:2018, bellec:2018, neftci:2019}. 
The loss function is then
\begin{align}
l_\rho(W_{\text{rec}}, W_{\text{in}}) = \lvert \rho - f(\chi(\pi(s))) \rvert^2_2,
\end{align}
where $\chi$ is defined as before (\ref{eq:exp_correlations}) by the spin expectation values and correlations 
in equations (\ref{eq:spin_average_1}-\ref{eq:spin_average_2}). 

\begin{figure}
\includegraphics{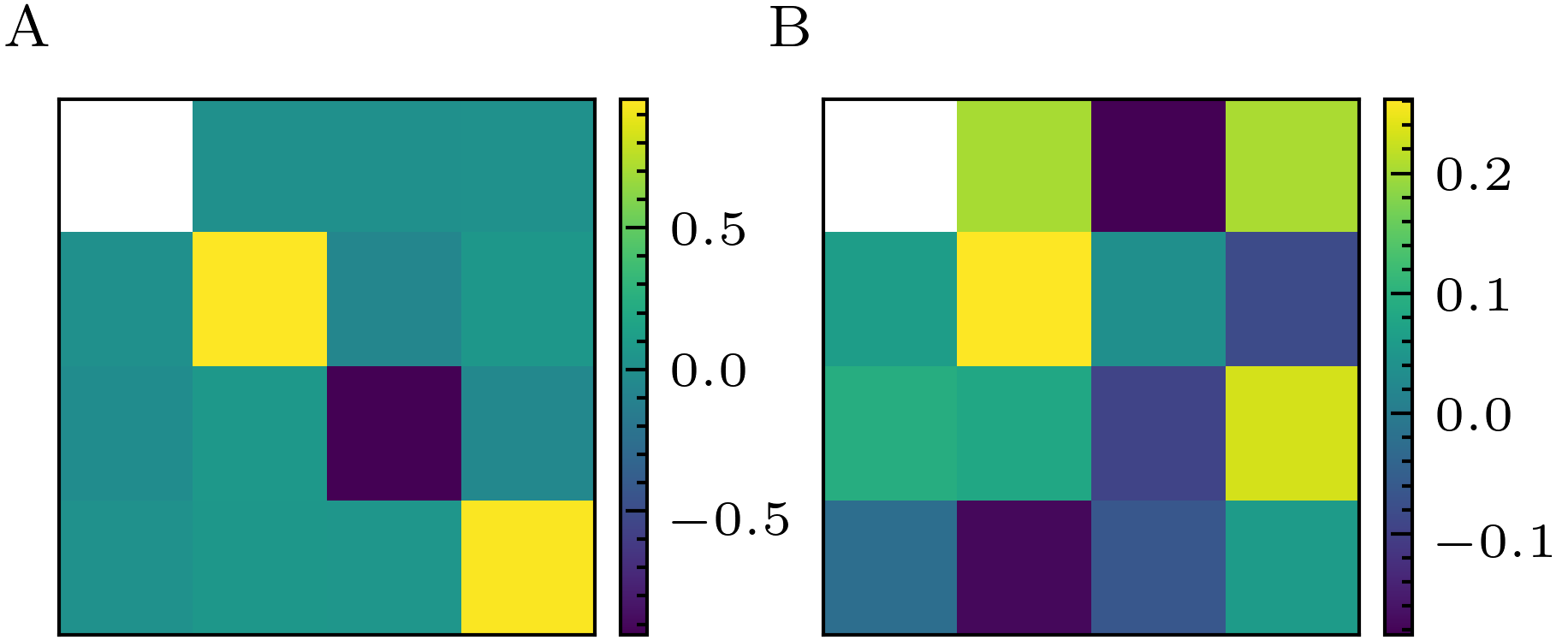}
\caption{Spin expectation and correlation matrices $\chi$ corresponding
to the pure state $\psi_+$ (A) and $\rho$ (B) respectively.}
\label{fig:spin_expectation_and_correlation}
\end{figure}

\begin{figure}
\includegraphics{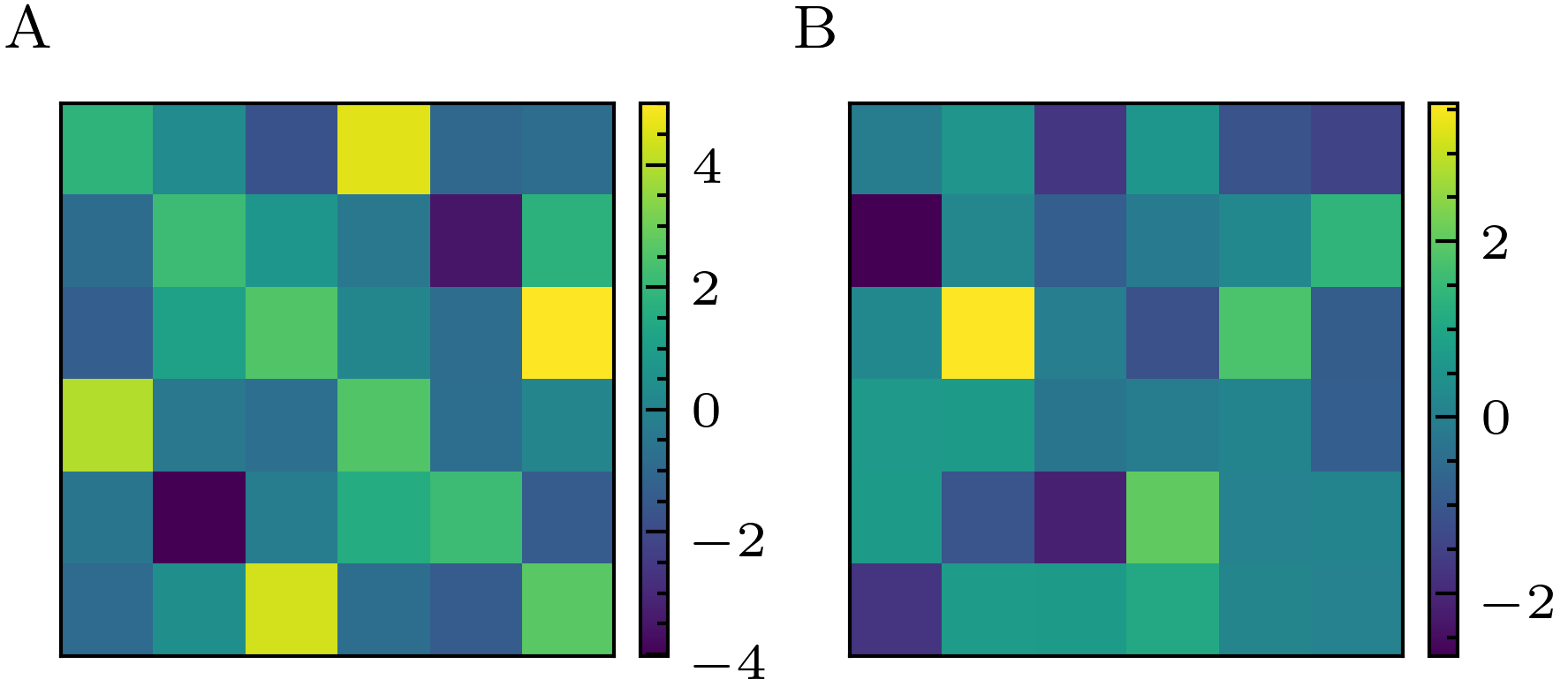}
\caption{Final recurrent weight matrices $W_{\text{rec}}$ or the $6$ recurrently
connected neurons, whose refractory state corresponds to the spins $\spin{i}{k}$
of the pure state $\psi_+$ (A) and $\rho$ (B) respectively.}
\label{fig:weights_for_densities}
\end{figure}

In figures \ref{figure:spike_loss}, \ref{fig:spin_expectation_and_correlation} we show the result of the optimization process
for the spin expectation and correlation matrices $\chi$. We also indicate in figure \ref{fig:weights_for_densities}
 a view of the resulting recurrent weight matrix $W_{\text{rec}}$ restricted to the
spins $s^{(i)}_k$. The membrane threshold is set to one, $v_{\text{th}} = 1,$ and 
the leak and reset potentials to zero, $v_l = v_r = 0$. We integrate for $T = 10^5$ timesteps with an integration step of $\text{dt} = 1 \text{ms}$.  
We choose an input dimension of $m = 128$ and consider $n = 64$ recurrently connected LIF neurons. 
We set $t_{\text{refrac}} = \text{dt}$ (this eliminates the need to take the equations for the refractory state into account) 
and draw the input spikes with poisson frequency $\lambda = 700 \text{Hz}$. 
The learning rate of the gradient descent starts at $\eta = 10$ and is exponentially 
decreased with a decay constant of $1/100$. The numerical implementation was done
in JAX \cite{jax:2018} using simple forward Euler integration. 

For the result of the optimization we plot the fidelity, which is a common measure to judge
how well a given quantum state is approximated. The fidelity is defined by $F(\rho, \sigma) = (\text{tr} \sqrt{\sqrt{\rho} \sigma \sqrt{\rho}})^2$. Instead 
of optimizing for the fidelity directly we minimize the square of the Frobenius norm $\lvert \rho - \sigma \rvert_2^2$.
By the Fuchs–van de Graaf inequalities we know that 
${\displaystyle 1-{\sqrt {F(\rho ,\sigma )}}\leq \frac{1}{2} \lvert \rho - \sigma \rvert_{1} \leq {\sqrt {1-F(\rho ,\sigma )}}\,}.$ 
Since the trace norm is in turn bounded by the Frobenius norm
the fidelity approaches one  as the square of the Frobenius norm goes to zero. 
Computing the fidelity directly is computationally more expensive and also 
has the added disadvantage that it isn't real valued for arbitrary complex matrices $\rho, \sigma$.

While the learning for the entangled pure state takes somewhat longer than for the randomly chosen state, it is clear that
after a reasonable learning time the neuronal dynamics has adapted to represent the quantum density matrix
with acceptable fidelity. Combined with the learning of the unitary quantum gates above one concludes that this type of neural
network can learn unitary transformations for two qubit quantum systems. Details of an optimal learning algorithm for spiking neural networks remain to be worked out. 

\section{Generalisations to many qubits}
So far we focussed on the case of two qubits. The correlation map can be extended to
$n$-qubits. Using generators
\begin{equation}
L_{\mu_1 \ldots \mu_n} = \bigotimes^n_{i=1} \tau_{\mu_i} \quad \mu_i = 0 \ldots 3,
\end{equation}
we write the density matrix as
\begin{equation}
\label{eq:ndim_density}
\rho = \frac{1}{2^n} \chi_{\mu_1 \ldots \mu_n} L_{\mu_1 \ldots \mu_n}.
\end{equation}
For the minimal correlation map the coefficients $\chi_{\mu_1 \ldots \mu_n}$ are determined by correlating $3 n$ spins as follows: 
Write $s^{(i)}_\mu(t) = (1, s^{(i)}_k(t))$ with $i = 1..n$, $k = 1..3$ and $\mu = 0 \ldots 3$. 
Define 
\begin{equation}
\label{eq:ndim_spin}
\sigma_{\mu_1 \ldots \mu_n}(t) = s^{(1)}_{\mu_1}(t) \ldots s^{(n)}_{\mu_n}(t),
\end{equation}
and
\begin{equation}
\label{eq:ndim_chi}
\chi_{\mu_1 \ldots \mu_n} = \frac{1}{T} \int_0^T \sigma_{\mu_1 \ldots \mu_n}(t) \text{dt}.
\end{equation}
This procedure involves up to $n$-point correlations. One may ask if the minimal
correlation map remains a complete bit-quantum map for three qubits. We can employ
the same methodology for $Q = 3$ as previously for $Q = 2$. Expectation values and
correlations are evaluated by time averaging for nine Ising spins and corresponding selected 
nine neurons. 

Time averaging of Ising spins representing some quantity above a threshhold ($s_j = 1$)
or below a threshhold ($s_j = -1$) is comparatively economical in this respect. It is sufficient to measure
for all Ising spins $s_j$ the times above or below threshold. With $t^-_j = T - t^+_j$ one can use identities
of the type
\begin{align}
\langle s_j \rangle &= \frac{t^+_j - t^-_j}{T} = \frac{2 t^+_j}{T} - 1,\\
\langle s_j s_i \rangle &= \frac{2 (t^{++}_{ji} + t^{--}_{ji})}{T} - 1,\\
\langle s_k s_j s_i \rangle &= \frac{2 (t^{+++}_{jik} + t^{+--}_{jik} + t^{-+-}_{jik} + t^{--+}_{jik})}{T} - 1,
\end{align}
where  $t^{++}_{ji}$ is the time when both $s_j$ and $s_i$ are above threshhold and so on.

We find that for three qubits an obstruction prevents the minimal correlation map
to be complete. There are valid quantum density matrices for which no 
classical correlation functions can be realized that obey eqs \eqref{eq:ndim_density}, \eqref{eq:ndim_spin} \eqref{eq:ndim_chi}.
For this purpose we concentrate on the GHZ-states
\begin{equation}\label{eq:ghz3_state}
\psi = \frac{1}{\sqrt{2}} (\rvert + + + \rangle + \varepsilon \rvert --- \rangle),\quad \lvert \varepsilon \rvert = 1.
\end{equation}
The only non-vanishing elements of the corresponding pure state density matrix $\rho_{\text{ghz}}$
are 
\begin{align}
\rho_{+++,+++} = \rho_{---,---} = \frac{1}{2}, \nonumber \\
\rho_{+++,---} = \frac{\varepsilon^*}{2}, \rho_{---,+++} = \frac{\varepsilon}{2}.
\end{align}
The elements $\sigma_{\mu_1 \mu_2 \mu_3}$ are given by
\begin{align}
\rho &= \frac{1}{2} \sigma_{\mu_1 \mu_2 \mu_3} \tau_{\mu_1} \otimes \tau_{\mu_2} \otimes \tau_{\mu_3} \nonumber \\
     &= \frac{1}{16} \{ (1 + \tau_3) \otimes (1 + \tau_3) \otimes (1+\tau_3) \nonumber \\
     &+ (1 - \tau_3) \otimes (1 - \tau_3) \otimes (1-\tau_3) \nonumber \\
     &+ \varepsilon (\tau_1 - i \tau_2) \otimes (\tau_1 - i \tau_2) \otimes (\tau_1 - i \tau_2) \nonumber \\
     &+ \varepsilon^* (\tau_1 + i \tau_2)\otimes (\tau_1 + i \tau_2) \otimes (\tau_1 + i \tau_2),
\end{align}
which results in non-zero values for 
\begin{align}
\label{eq:ghz3_expectation_values_1}
\sigma_{000} &= \sigma_{330} = \sigma_{303} = \sigma_{033} = 1, \\
\label{eq:ghz3_expectation_values_2}
\sigma_{111} &= -\sigma_{221} = -\sigma_{122} = - \sigma_{212} = \frac{1}{2} (\varepsilon + \varepsilon^*),\\
\label{eq:ghz3_expectation_values_3}
\sigma_{112} &= \sigma_{211} = \sigma_{121} = -\sigma_{222} = -\frac{i}{2} (\varepsilon -\varepsilon^*).
\end{align}

\begin{figure}
\includegraphics{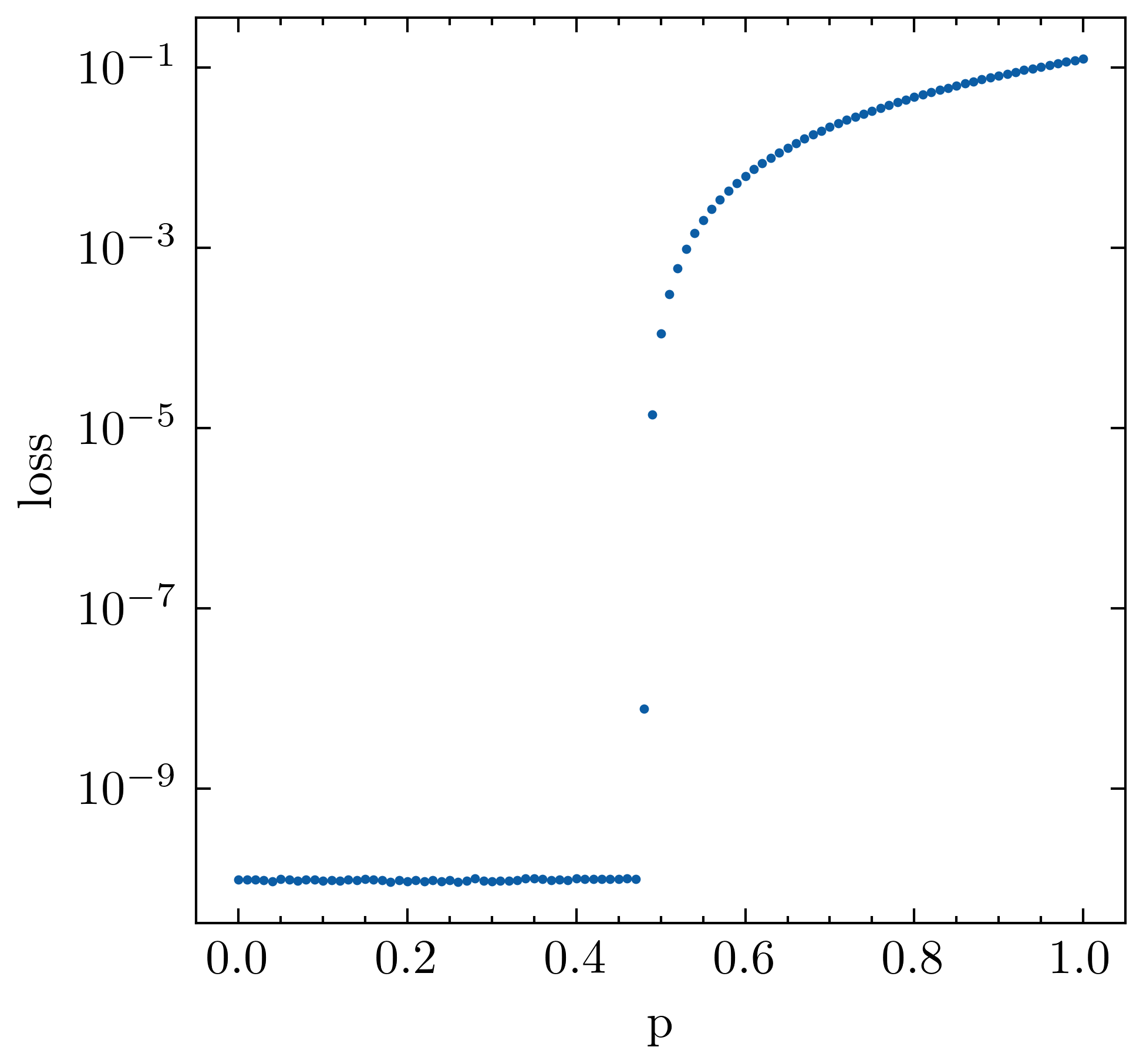}
\caption{\label{fig:loss_ghz3} Final loss after training up to $10^4$ epochs to
approximate $\rho(p) = p \rho_{\text{ghz}} + (1-p) \bar{\rho}$, where
$\rho$ is a randomly chosen density matrix and $\rho_{\text{ghz}}$ is 
the density matrix of the GHZ-state in dimension $3$.}
\end{figure}

\begin{figure}
\includegraphics{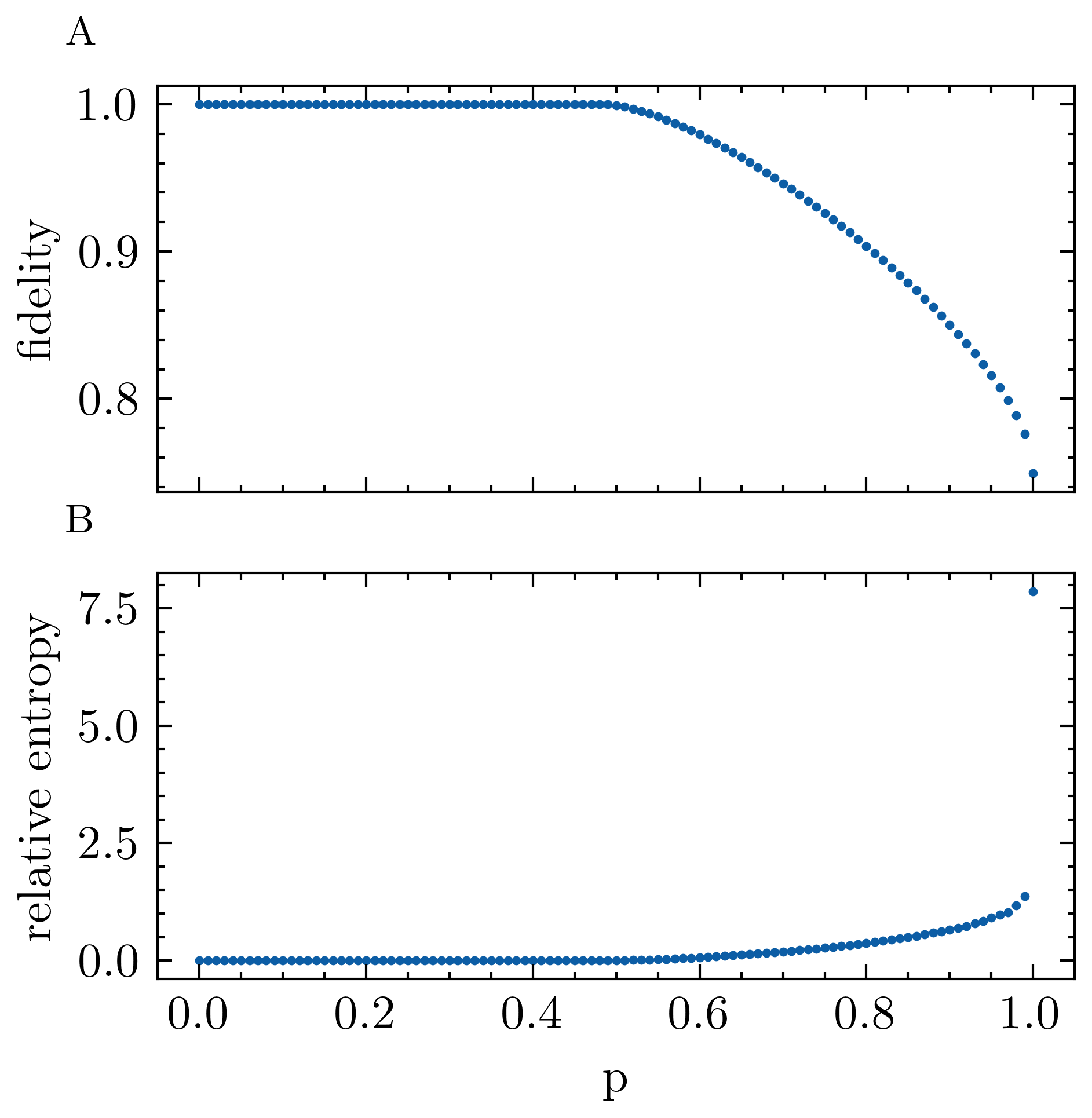}
\caption{Final fidelity (A) and relative entropy (B) 
after training up to $10^4$ epochs to approximate $\rho(p) = p \rho_{\text{ghz}} + (1-p) \bar{\rho}$, where
$\rho$ is a randomly chosen density matrix and $\rho_{\text{ghz}}$ is 
the density matrix of the GHZ-state in dimension $3$.}
\label{fig:re_fid_ghz_3}
\end{figure}

For randomly chosen density matrices one finds for most cases suitable probability distributions
that realise them by eq. \eqref{eq:ndim_density}, \eqref{eq:ndim_spin}, \eqref{eq:ndim_chi}. If we look instead at a special
class of density matrices
\begin{equation}
\rho = p \rho_{\text{ghz}} + (1-p) \bar{\rho},
\end{equation}
where $\bar{\rho}$ is chosen randomly and $0 \leq p \leq 1$, 
for large enough $p$ no such probability distribution can be found anymore.
This is demonstrated in Fig. \ref{fig:loss_ghz3}, where we plot the
loss after training for up to $10^4$ epochs against $p$.

This is reflected in the fidelity and relative entropy for a comparison of the finally optimized
matrix and the true $GHZ$-density matrix, shown in Fig. \ref{fig:re_fid_ghz_3}.
Again we observe a clear insufficiency of the minimal correlation map for a reproduction
of the density matrices close to the GHZ-state. This insufficiency sets
in rather sharply at a certain value of $p$. 

One can understand the obstruction analytically. For this reason we consder the GHZ-density
matrix with $\varepsilon = 1$. The expectation values \eqref{eq:ghz3_expectation_values_1}, \eqref{eq:ghz3_expectation_values_2}, \eqref{eq:ghz3_expectation_values_3}
can be realized by a a factorizing probability distribution,
\begin{equation}
\label{eq:p_direct_product_ghz_3}
p = p_1[s^{(i)}_3] \: p_2[s^{(i)}_1, s^{(i)}_2].
\end{equation}
The first factor $p_1$ has to realise the correlations
\begin{align}
\langle s^{(1)}_3 s^{(2)}_3 \rangle &= \langle s^{(1)}_3 s^{(3)}_3\rangle = \langle s^{(2)}_3 s^{(3)}_3\rangle = 1,\nonumber \\
\langle s^{(i)}_3 \rangle &= 0, \quad \langle s^{(1)}_3 s^{(2)}_3 s^{(3)}_3 \rangle = 0.
\end{align}
This is achieved by the probabilities
\begin{equation}
p_{+++} = p_{---} = \frac{1}{2}.
\end{equation}
The second factor $p_2$ depends on the Ising spins $s^{(i)}_1$ and $s^{(i)}_2$ 
and has to realise the correlations
\begin{equation}
\label{eq:three_point_spin_ghz_3_0}
\langle s^{(1)}_1 s^{(2)}_2 s^{(3)}_2 \rangle = \langle s^{(1)}_2 s^{(2)}_1 s^{(3)}_2\rangle = \langle   s^{(1)}_2 s^{(2)}_2 s^{(3)}_1 \rangle = -1,
\end{equation}
and
\begin{equation}
\label{eq:three_point_spin_ghz_3}
\langle s^{(1)}_1 s^{(2)}_1 s^{(3)}_1 \rangle = 1.
\end{equation}

This is not possible for a classical statistical setting. Eq. \eqref{eq:three_point_spin_ghz_3}
requires that for the three spins $(s^{(1)}_1 s^{(2)}_1 s^{(3)}_1)$ only the 
configurations $(+++), (+--), (-+-), (--+)$ can have a non-zero probability.
For the case of the configuration $(+++)$ for $s^{(i)}_1$ eq. \eqref{eq:three_point_spin_ghz_3_0}
requires that the only configurations for $s^{(i)}_2$ that can have non-vanishing
probabilities must obey $s^{(2)}_2  s^{(3)}_2 = -1, s^{(1)}_2  s^{(2)}_2 = -1, s^{(1)}_2  s^{(3)}_2 = -1.$
Otherwise one of the three three-point functions \eqref{eq:three_point_spin_ghz_3_0} would be larger
than $-1$. If $s^{(1)}_2$ and $s^{(2)}_2$ have opposite signs, and $s^{(1)}_2$ and $s^{(3)}_2$ 
have opposite signs, one infers that $s^{(2)}_2$ and $s^{(3)}_2$ have the same sign,
in contradiction to $s^{(2)}_2 s^{(3)}_2 = -1.$ One concludes that the probability for
$(s^{(1)}_1 s^{(2)}_1 s^{(3)}_1 ) = (+++)$ must be zero. Similar chains of arguments
show that the three other configurations for $\spin{i}{1}$, namely $(+--), (-+-), (--+),$ cannot
have a non-zero probability either. In consequence, eq. \eqref{eq:three_point_spin_ghz_3}
cannot be obeyed. There is no probability distribution $p_2$ that can generate the set
of correlations \eqref{eq:three_point_spin_ghz_3}, \eqref{eq:three_point_spin_ghz_3_0}. This argument
generalises to probability distributions that are not of the direct product form
\eqref{eq:p_direct_product_ghz_3}. We conclude that for three qubits the minimal
correlation map is not complete.

One may envisage an extended correlation map with additonal $27$ spins ($\spin{12}{k_1 k_2}, \spin{13}{k_1 k_3}, \spin{23}{k_2 k_3}$),
$k_i = 1\ldots 3.$ The spins $\sigma_{\mu_1 \mu_2 \mu_3}$ with precisely one index zero are given by
\begin{align}
\sigma_{k_1 k_2 0} = \spin{12}{k_1 k_2}, \quad \sigma_{k_1 0 k_2} = \spin{13}{k_1 k_3}, \quad \sigma_{0 k_2 k_3} = \spin{23}{k_2 k_3},
\end{align}
instead of equation \eqref{eq:ndim_spin} for the minimal correlation map. Eq. \eqref{eq:ndim_chi} continues
to hold for all $\chi_{\mu_1 \mu_2 \mu_3}$ with one, two or three indices equal to zero. The
coefficients of the density matrix with only non-zero indices are correlations of "pair spins"
$\spin{ij}{kl}$ and "single spins" $\spin{i}{k}$,
\begin{equation}
\label{eq:higher_spins}
\chi_{k_1 k_2 k_3} = \langle \spin{12}{k_1 k_2} \spin{3}{k_3}\rangle = \langle \spin{13}{k_1 k_3} \spin{2}{k_2} \rangle = \langle \spin{23}{k_2 k_3} \spin{1}{k_1} \rangle.
\end{equation}
Eq. \eqref{eq:higher_spins} requires new quantum constraints on the classical probability distribution
since all three correlations have to be the same. In the presence of this constraint knowledge of one set 
of correlations, say $\langle \spin{12}{k_1 k_2} \spin{3}{k_3}\rangle$, gives access to
$\langle \spin{13}{k_1 k_3} \spin{2}{k_2} \rangle$ and  $\langle \spin{23}{k_2 k_3} \spin{1}{k_1} \rangle$.
This feature is typical for quantum systems. 

The GHZ-state with $\varepsilon = 1$ can be realized by the extended correlation map. For an
explicit construction of a classical probability distribution realizing this state we choose probabilities
which only differ from zero if
\begin{align}
\spin{23}{11} = -\spin{23}{22} = \spin{1}{1},\quad \spin{23}{12} = \spin{23}{21} = - \spin{1}{2},\\
\spin{13}{11} = -\spin{13}{22} = \spin{2}{1},\quad \spin{13}{12} = \spin{13}{21} = - \spin{2}{2},\\
\spin{12}{11} = -\spin{12}{22} = \spin{3}{1},\quad \spin{12}{12} = \spin{12}{21} = - \spin{3}{2}.
\end{align}
This guarantees the relations \eqref{eq:ghz3_expectation_values_2},
\begin{equation}
\label{eq:three_ghz3}
\sigma_{111} = -\sigma_{221} = -\sigma_{122} = - \sigma_{212} = 1,
\end{equation}
together with the constraint \eqref{eq:higher_spins} for the four quantities 
in \eqref{eq:three_ghz3}. The correlations in eq. \eqref{eq:ghz3_expectation_values_1}
are obeyed if non-zero probabilities occur only for
\begin{equation}
\label{eq:two_ghz3}
\spin{12}{33} = \spin{13}{33} = \spin{23}{33} = 1.
\end{equation}
With eqs. \eqref{eq:three_ghz3}, \eqref{eq:two_ghz3} the configurations with 
nonzero probabilities can be characterised by the values of the $21$ spins
$\bar{s}_a = (\spin{i}{k}, \spin{ij}{13}, \spin{ij}{23}, \spin{ij}{31}, \spin{ij}{32}).$
With the conditions
\begin{equation}
\label{eq:higher_spin_relation}
\langle \bar{s}_a \rangle = 0,\quad \langle \bar{s}_a \bar{s}_b \rangle = 0,
\end{equation}
all other coefficients except $\sigma_{111}$, $\sigma_{221}$, $\sigma_{122}$, $\sigma_{212}$, $\sigma_{330}$, $\sigma_{303}$, $\sigma_{033}$ and $\sigma_{000}$ vanish. 
The relation \eqref{eq:higher_spin_relation} can be realized by equipartition for the configurations
of spins $\bar{s}_a.$ This probability distribution realises the GHZ-state.

One can generalise the extended correlation map to a higher number of qubits $Q > 3.$ One chooses independent
spins for all $\sigma_{\mu_1 \ldots \mu_Q}$ with one or two "space"-indices $k, l$ taking values $k, l = 1\ldots 3$
and all other $\mu_i$ equal zero. These are $\frac{9}{2} Q^2 - \frac{3}{2} Q$ Ising spins, such that the 
number of classical spins grows quadratically with the number of qubits. Coefficients $\chi_{\mu_1 \ldots \mu_Q}$
with one or two space indices are given by the expectation values of the corresponding $\sigma_{\mu_1 \ldots \mu_Q}$. 
Coefficients with three space indices are correlations of one pair spin and one simple spin, with quantum constraint
\eqref{eq:higher_spins} extended correspondingly. For four space indices of $\chi_{\mu_1 \ldots \mu_Q}$
one takes correlations of two pair spins with corresponding quantum constraints. Five spin 
indices are realized by three point functions of two pair spins and one single spin, and so forth.
There is never more than one single spin in the correlations. It is not known
if this extended correlation map is a complete bit-quantum map, or if new obstructions
arise for a certain number of qubits.

\section{Incomplete probabilistic information}
The number $2^{2Q}$ of elements of the density matrix grows very rapidly with an
increasing number $Q$ of qubits. This issue is common to all approaches which use at every step of the
computation the full information about the quantum state. We find it unlikely that computations
by real quantum systems or artificial neurons need the full information contained in the density
matrix $\rho$. It becomes then an important task to find out which part of the probabilistic information
is involved for practical questions.

As an example we consider the search for the ground state energy of the one-dimensional quantum Ising model
with Hamiltonian
\begin{equation}
H =-J\sum _{i=1,\ldots ,n} \tau_{3}^{(i)}\tau_{3}^{(i+1)}-h\sum _{i}\tau_{1}^{(i)},
\end{equation}
one of the benchmark tasks in \cite{carleo:2016}. The goal is to find a 
density matrix $\rho$, such that $\text{tr}(H \rho)$ is minimized.
This requires only a small subset of the $\chi_{\mu_1 \ldots \mu_n},$
namely those $n$ two-point functions 
for which $\mu_i = \mu_{i+1} = 3$ and all other $\mu_k = 0$, as well as $n$ expectation
values for which  $\mu_i = 1$ and all other $\mu_k = 0$. For the minimal correlation map
the first set of $\chi$ is given by two point correlation functions. These two point
correlation functions cannot take arbitrary values, however. The positivity of the
density matrix imposes "quantum constraints" \cite{wetterich:2018, wetterich2020probabilistic}
that these correlation functions have to obey. Due to these constraints the minimal value 
of $\langle H \rangle$ is higher than for unconstrained correlation functions. For the 
extended correlation map also the first set of $\chi$ is given by expectation values. Again,
quantum constraints restrict the possible values.

The explicit use of all quantum constraints seems a difficult task for a high number of qubits. It 
may, however, be sufficient to impose only part of the quantum constraints for obtaining already a 
reasonable approximation to the minimal value of $\langle H \rangle$. These partial constraints
could then be associated with the information that is relevant for a given problem, while additional
information concerning the full set of quantum constraints may be discarded. A probabilistic view
on expectation values and correlations may help to focus on the relevant information needed for a given
quantum problem.

The computation of the expectation values $\chi_{\mu_1 \ldots \mu_n}$ involves purely classical probabilistic settings,
as neurons firing in biological or artificial systems. No low temperature or a high degree of isolation of small
subsystems is needed for such a realization of quantum features. Our findings are an example how quantum evolution
can be realized by probabilistic classical systems \cite{wetterich:2018information,wetterich:2018quantum,wetterich2020probabilistic}.

This work was supported by the EU Horizon 2020 framework program under grant agreement 720270 (Human Brain Project), the Heidelberg Graduate School for Fundamental Physics (HGSFP)
and the DFG Collaborative Research Centre SFB 1225 (ISOQUANT).

\bibliography{bibliography}

\begin{thebibliography}{22}%
\makeatletter
\providecommand \@ifxundefined [1]{%
 \@ifx{#1\undefined}
}%
\providecommand \@ifnum [1]{%
 \ifnum #1\expandafter \@firstoftwo
 \else \expandafter \@secondoftwo
 \fi
}%
\providecommand \@ifx [1]{%
 \ifx #1\expandafter \@firstoftwo
 \else \expandafter \@secondoftwo
 \fi
}%
\providecommand \natexlab [1]{#1}%
\providecommand \enquote  [1]{``#1''}%
\providecommand \bibnamefont  [1]{#1}%
\providecommand \bibfnamefont [1]{#1}%
\providecommand \citenamefont [1]{#1}%
\providecommand \href@noop [0]{\@secondoftwo}%
\providecommand \href [0]{\begingroup \@sanitize@url \@href}%
\providecommand \@href[1]{\@@startlink{#1}\@@href}%
\providecommand \@@href[1]{\endgroup#1\@@endlink}%
\providecommand \@sanitize@url [0]{\catcode `\\12\catcode `\$12\catcode
  `\&12\catcode `\#12\catcode `\^12\catcode `\_12\catcode `\%12\relax}%
\providecommand \@@startlink[1]{}%
\providecommand \@@endlink[0]{}%
\providecommand \url  [0]{\begingroup\@sanitize@url \@url }%
\providecommand \@url [1]{\endgroup\@href {#1}{\urlprefix }}%
\providecommand \urlprefix  [0]{URL }%
\providecommand \Eprint [0]{\href }%
\providecommand \doibase [0]{https://doi.org/}%
\providecommand \selectlanguage [0]{\@gobble}%
\providecommand \bibinfo  [0]{\@secondoftwo}%
\providecommand \bibfield  [0]{\@secondoftwo}%
\providecommand \translation [1]{[#1]}%
\providecommand \BibitemOpen [0]{}%
\providecommand \bibitemStop [0]{}%
\providecommand \bibitemNoStop [0]{.\EOS\space}%
\providecommand \EOS [0]{\spacefactor3000\relax}%
\providecommand \BibitemShut  [1]{\csname bibitem#1\endcsname}%
\let\auto@bib@innerbib\@empty
\bibitem [{\citenamefont {Wetterich}(2019)}]{wetterich:2018}%
  \BibitemOpen
  \bibfield  {author} {\bibinfo {author} {\bibfnamefont {C.}~\bibnamefont
  {Wetterich}},\ }\bibfield  {title} {\bibinfo {title} {{Quantum computing with
  classical bits}},\ }\href {https://doi.org/10.1016/j.nuclphysb.2019.114776}
  {\bibfield  {journal} {\bibinfo  {journal} {Nucl. Phys. B}\ }\textbf
  {\bibinfo {volume} {948}},\ \bibinfo {pages} {114776} (\bibinfo {year}
  {2019})},\ \Eprint {https://arxiv.org/abs/1806.05960} {arXiv:1806.05960
  [quant-ph]} \BibitemShut {NoStop}%
\bibitem [{\citenamefont {Pehle}\ \emph {et~al.}(2018)\citenamefont {Pehle},
  \citenamefont {Meier}, \citenamefont {Oberthaler},\ and\ \citenamefont
  {Wetterich}}]{pehle:2018}%
  \BibitemOpen
  \bibfield  {author} {\bibinfo {author} {\bibfnamefont {C.}~\bibnamefont
  {Pehle}}, \bibinfo {author} {\bibfnamefont {K.}~\bibnamefont {Meier}},
  \bibinfo {author} {\bibfnamefont {M.}~\bibnamefont {Oberthaler}},\ and\
  \bibinfo {author} {\bibfnamefont {C.}~\bibnamefont {Wetterich}},\ }\bibfield
  {title} {\bibinfo {title} {Emulating quantum computation with artificial
  neural networks},\ }\href@noop {} {\bibfield  {journal} {\bibinfo  {journal}
  {arXiv preprint arXiv:1810.10335}\ } (\bibinfo {year} {2018})}\BibitemShut
  {NoStop}%
\bibitem [{\citenamefont {{Carleo}}\ and\ \citenamefont
  {{Troyer}}(2017)}]{carleo:2016}%
  \BibitemOpen
  \bibfield  {author} {\bibinfo {author} {\bibfnamefont {G.}~\bibnamefont
  {{Carleo}}}\ and\ \bibinfo {author} {\bibfnamefont {M.}~\bibnamefont
  {{Troyer}}},\ }\bibfield  {title} {\bibinfo {title} {{Solving the quantum
  many-body problem with artificial neural networks}},\ }\href
  {https://doi.org/10.1126/science.aag2302} {\bibfield  {journal} {\bibinfo
  {journal} {Science}\ }\textbf {\bibinfo {volume} {355}},\ \bibinfo {pages}
  {602} (\bibinfo {year} {2017})},\ \Eprint {https://arxiv.org/abs/1606.02318}
  {arXiv:1606.02318 [cond-mat.dis-nn]} \BibitemShut {NoStop}%
\bibitem [{\citenamefont {Torlai}\ \emph {et~al.}(2018)\citenamefont {Torlai},
  \citenamefont {Mazzola}, \citenamefont {Carrasquilla}, \citenamefont
  {Troyer}, \citenamefont {Melko},\ and\ \citenamefont {Carleo}}]{torlai:2018}%
  \BibitemOpen
  \bibfield  {author} {\bibinfo {author} {\bibfnamefont {G.}~\bibnamefont
  {Torlai}}, \bibinfo {author} {\bibfnamefont {G.}~\bibnamefont {Mazzola}},
  \bibinfo {author} {\bibfnamefont {J.}~\bibnamefont {Carrasquilla}}, \bibinfo
  {author} {\bibfnamefont {M.}~\bibnamefont {Troyer}}, \bibinfo {author}
  {\bibfnamefont {R.}~\bibnamefont {Melko}},\ and\ \bibinfo {author}
  {\bibfnamefont {G.}~\bibnamefont {Carleo}},\ }\bibfield  {title} {\bibinfo
  {title} {Neural-network quantum state tomography},\ }\href
  {https://doi.org/10.1038/s41567-018-0048-5} {\bibfield  {journal} {\bibinfo
  {journal} {Nature Physics}\ }\textbf {\bibinfo {volume} {14}},\ \bibinfo
  {pages} {447} (\bibinfo {year} {2018})}\BibitemShut {NoStop}%
\bibitem [{\citenamefont {Sharir}\ \emph {et~al.}(2020)\citenamefont {Sharir},
  \citenamefont {Levine}, \citenamefont {Wies}, \citenamefont {Carleo},\ and\
  \citenamefont {Shashua}}]{sharir:2020}%
  \BibitemOpen
  \bibfield  {author} {\bibinfo {author} {\bibfnamefont {O.}~\bibnamefont
  {Sharir}}, \bibinfo {author} {\bibfnamefont {Y.}~\bibnamefont {Levine}},
  \bibinfo {author} {\bibfnamefont {N.}~\bibnamefont {Wies}}, \bibinfo {author}
  {\bibfnamefont {G.}~\bibnamefont {Carleo}},\ and\ \bibinfo {author}
  {\bibfnamefont {A.}~\bibnamefont {Shashua}},\ }\bibfield  {title} {\bibinfo
  {title} {Deep autoregressive models for the efficient variational simulation
  of many-body quantum systems},\ }\href
  {https://doi.org/10.1103/PhysRevLett.124.020503} {\bibfield  {journal}
  {\bibinfo  {journal} {Phys. Rev. Lett.}\ }\textbf {\bibinfo {volume} {124}},\
  \bibinfo {pages} {020503} (\bibinfo {year} {2020})}\BibitemShut {NoStop}%
\bibitem [{\citenamefont {Broughton}\ \emph {et~al.}(2020)\citenamefont
  {Broughton}, \citenamefont {Verdon}, \citenamefont {McCourt}, \citenamefont
  {Martinez}, \citenamefont {Yoo}, \citenamefont {Isakov}, \citenamefont
  {Massey}, \citenamefont {Niu}, \citenamefont {Halavati}, \citenamefont
  {Peters}, \citenamefont {Leib}, \citenamefont {Skolik}, \citenamefont
  {Streif}, \citenamefont {Dollen}, \citenamefont {McClean}, \citenamefont
  {Boixo}, \citenamefont {Bacon}, \citenamefont {Ho}, \citenamefont {Neven},\
  and\ \citenamefont {Mohseni}}]{broughton2020tensorflow}%
  \BibitemOpen
  \bibfield  {author} {\bibinfo {author} {\bibfnamefont {M.}~\bibnamefont
  {Broughton}}, \bibinfo {author} {\bibfnamefont {G.}~\bibnamefont {Verdon}},
  \bibinfo {author} {\bibfnamefont {T.}~\bibnamefont {McCourt}}, \bibinfo
  {author} {\bibfnamefont {A.~J.}\ \bibnamefont {Martinez}}, \bibinfo {author}
  {\bibfnamefont {J.~H.}\ \bibnamefont {Yoo}}, \bibinfo {author} {\bibfnamefont
  {S.~V.}\ \bibnamefont {Isakov}}, \bibinfo {author} {\bibfnamefont
  {P.}~\bibnamefont {Massey}}, \bibinfo {author} {\bibfnamefont {M.~Y.}\
  \bibnamefont {Niu}}, \bibinfo {author} {\bibfnamefont {R.}~\bibnamefont
  {Halavati}}, \bibinfo {author} {\bibfnamefont {E.}~\bibnamefont {Peters}},
  \bibinfo {author} {\bibfnamefont {M.}~\bibnamefont {Leib}}, \bibinfo {author}
  {\bibfnamefont {A.}~\bibnamefont {Skolik}}, \bibinfo {author} {\bibfnamefont
  {M.}~\bibnamefont {Streif}}, \bibinfo {author} {\bibfnamefont {D.~V.}\
  \bibnamefont {Dollen}}, \bibinfo {author} {\bibfnamefont {J.~R.}\
  \bibnamefont {McClean}}, \bibinfo {author} {\bibfnamefont {S.}~\bibnamefont
  {Boixo}}, \bibinfo {author} {\bibfnamefont {D.}~\bibnamefont {Bacon}},
  \bibinfo {author} {\bibfnamefont {A.~K.}\ \bibnamefont {Ho}}, \bibinfo
  {author} {\bibfnamefont {H.}~\bibnamefont {Neven}},\ and\ \bibinfo {author}
  {\bibfnamefont {M.}~\bibnamefont {Mohseni}},\ }\href@noop {} {\bibinfo
  {title} {Tensorflow quantum: A software framework for quantum machine
  learning}} (\bibinfo {year} {2020}),\ \Eprint
  {https://arxiv.org/abs/2003.02989} {arXiv:2003.02989 [quant-ph]} \BibitemShut
  {NoStop}%
\bibitem [{\citenamefont {Hopfield}(1982)}]{hopfield:1982}%
  \BibitemOpen
  \bibfield  {author} {\bibinfo {author} {\bibfnamefont {J.~J.}\ \bibnamefont
  {Hopfield}},\ }\bibfield  {title} {\bibinfo {title} {Neural networks and
  physical systems with emergent collective computational abilities},\
  }\href@noop {} {\bibfield  {journal} {\bibinfo  {journal} {Proceedings of the
  national academy of sciences}\ }\textbf {\bibinfo {volume} {79}},\ \bibinfo
  {pages} {2554} (\bibinfo {year} {1982})}\BibitemShut {NoStop}%
\bibitem [{\citenamefont {James}\ \emph {et~al.}(2001)\citenamefont {James},
  \citenamefont {Kwiat}, \citenamefont {Munro},\ and\ \citenamefont
  {White}}]{james:2001}%
  \BibitemOpen
  \bibfield  {author} {\bibinfo {author} {\bibfnamefont {D.~F.~V.}\
  \bibnamefont {James}}, \bibinfo {author} {\bibfnamefont {P.~G.}\ \bibnamefont
  {Kwiat}}, \bibinfo {author} {\bibfnamefont {W.~J.}\ \bibnamefont {Munro}},\
  and\ \bibinfo {author} {\bibfnamefont {A.~G.}\ \bibnamefont {White}},\
  }\bibfield  {title} {\bibinfo {title} {Measurement of qubits},\ }\href
  {https://doi.org/10.1103/PhysRevA.64.052312} {\bibfield  {journal} {\bibinfo
  {journal} {Phys. Rev. A}\ }\textbf {\bibinfo {volume} {64}},\ \bibinfo
  {pages} {052312} (\bibinfo {year} {2001})}\BibitemShut {NoStop}%
\bibitem [{\citenamefont {Paris}\ and\ \citenamefont
  {Rehacek}(2004)}]{paris:2004}%
  \BibitemOpen
  \bibfield  {author} {\bibinfo {author} {\bibfnamefont {M.}~\bibnamefont
  {Paris}}\ and\ \bibinfo {author} {\bibfnamefont {J.}~\bibnamefont
  {Rehacek}},\ }\href@noop {} {\emph {\bibinfo {title} {Quantum state
  estimation}}},\ Vol.\ \bibinfo {volume} {649}\ (\bibinfo  {publisher}
  {Springer Science \& Business Media},\ \bibinfo {year} {2004})\BibitemShut
  {NoStop}%
\bibitem [{\citenamefont {Kiesel}(2007)}]{kiesel:2007}%
  \BibitemOpen
  \bibfield  {author} {\bibinfo {author} {\bibfnamefont {N.}~\bibnamefont
  {Kiesel}},\ }\emph {\bibinfo {title} {Experiments on multiphoton
  entanglement}},\ \href@noop {} {Ph.D. thesis},\ \bibinfo  {school} {lmu}
  (\bibinfo {year} {2007})\BibitemShut {NoStop}%
\bibitem [{\citenamefont {Schneidman}\ \emph {et~al.}(2006)\citenamefont
  {Schneidman}, \citenamefont {Berry}, \citenamefont {Segev},\ and\
  \citenamefont {Bialek}}]{schneidman:2006}%
  \BibitemOpen
  \bibfield  {author} {\bibinfo {author} {\bibfnamefont {E.}~\bibnamefont
  {Schneidman}}, \bibinfo {author} {\bibfnamefont {M.~J.}\ \bibnamefont
  {Berry}}, \bibinfo {author} {\bibfnamefont {R.}~\bibnamefont {Segev}},\ and\
  \bibinfo {author} {\bibfnamefont {W.}~\bibnamefont {Bialek}},\ }\bibfield
  {title} {\bibinfo {title} {Weak pairwise correlations imply strongly
  correlated network states in a neural population},\ }\href@noop {} {\bibfield
   {journal} {\bibinfo  {journal} {Nature}\ }\textbf {\bibinfo {volume}
  {440}},\ \bibinfo {pages} {1007} (\bibinfo {year} {2006})}\BibitemShut
  {NoStop}%
\bibitem [{\citenamefont {Ohiorhenuan}\ \emph {et~al.}(2010)\citenamefont
  {Ohiorhenuan}, \citenamefont {Mechler}, \citenamefont {Purpura},
  \citenamefont {Schmid}, \citenamefont {Hu},\ and\ \citenamefont
  {Victor}}]{ohiorhenuan:2010}%
  \BibitemOpen
  \bibfield  {author} {\bibinfo {author} {\bibfnamefont {I.~E.}\ \bibnamefont
  {Ohiorhenuan}}, \bibinfo {author} {\bibfnamefont {F.}~\bibnamefont
  {Mechler}}, \bibinfo {author} {\bibfnamefont {K.~P.}\ \bibnamefont
  {Purpura}}, \bibinfo {author} {\bibfnamefont {A.~M.}\ \bibnamefont {Schmid}},
  \bibinfo {author} {\bibfnamefont {Q.}~\bibnamefont {Hu}},\ and\ \bibinfo
  {author} {\bibfnamefont {J.~D.}\ \bibnamefont {Victor}},\ }\bibfield  {title}
  {\bibinfo {title} {Sparse coding and high-order correlations in fine-scale
  cortical networks},\ }\href@noop {} {\bibfield  {journal} {\bibinfo
  {journal} {Nature}\ }\textbf {\bibinfo {volume} {466}},\ \bibinfo {pages}
  {617} (\bibinfo {year} {2010})}\BibitemShut {NoStop}%
\bibitem [{\citenamefont {Buesing}\ \emph {et~al.}(2011)\citenamefont
  {Buesing}, \citenamefont {Bill}, \citenamefont {Nessler},\ and\ \citenamefont
  {Maass}}]{buesing:2011}%
  \BibitemOpen
  \bibfield  {author} {\bibinfo {author} {\bibfnamefont {L.}~\bibnamefont
  {Buesing}}, \bibinfo {author} {\bibfnamefont {J.}~\bibnamefont {Bill}},
  \bibinfo {author} {\bibfnamefont {B.}~\bibnamefont {Nessler}},\ and\ \bibinfo
  {author} {\bibfnamefont {W.}~\bibnamefont {Maass}},\ }\bibfield  {title}
  {\bibinfo {title} {Neural dynamics as sampling: A model for stochastic
  computation in recurrent networks of spiking neurons},\ }\href
  {https://doi.org/10.1371/journal.pcbi.1002211} {\bibfield  {journal}
  {\bibinfo  {journal} {PLOS Computational Biology}\ }\textbf {\bibinfo
  {volume} {7}},\ \bibinfo {pages} {1} (\bibinfo {year} {2011})}\BibitemShut
  {NoStop}%
\bibitem [{\citenamefont {Petrovici}\ \emph {et~al.}(2016)\citenamefont
  {Petrovici}, \citenamefont {Bill}, \citenamefont {Bytschok}, \citenamefont
  {Schemmel},\ and\ \citenamefont {Meier}}]{petrovici:2016}%
  \BibitemOpen
  \bibfield  {author} {\bibinfo {author} {\bibfnamefont {M.~A.}\ \bibnamefont
  {Petrovici}}, \bibinfo {author} {\bibfnamefont {J.}~\bibnamefont {Bill}},
  \bibinfo {author} {\bibfnamefont {I.}~\bibnamefont {Bytschok}}, \bibinfo
  {author} {\bibfnamefont {J.}~\bibnamefont {Schemmel}},\ and\ \bibinfo
  {author} {\bibfnamefont {K.}~\bibnamefont {Meier}},\ }\bibfield  {title}
  {\bibinfo {title} {Stochastic inference with spiking neurons in the
  high-conductance state},\ }\href {https://doi.org/10.1103/PhysRevE.94.042312}
  {\bibfield  {journal} {\bibinfo  {journal} {Phys. Rev. E}\ }\textbf {\bibinfo
  {volume} {94}},\ \bibinfo {pages} {042312} (\bibinfo {year}
  {2016})}\BibitemShut {NoStop}%
\bibitem [{\citenamefont {Zenke}\ and\ \citenamefont
  {Ganguli}(2018)}]{zenke:2018}%
  \BibitemOpen
  \bibfield  {author} {\bibinfo {author} {\bibfnamefont {F.}~\bibnamefont
  {Zenke}}\ and\ \bibinfo {author} {\bibfnamefont {S.}~\bibnamefont
  {Ganguli}},\ }\bibfield  {title} {\bibinfo {title} {Superspike: Supervised
  learning in multilayer spiking neural networks},\ }\href@noop {} {\bibfield
  {journal} {\bibinfo  {journal} {Neural computation}\ }\textbf {\bibinfo
  {volume} {30}},\ \bibinfo {pages} {1514} (\bibinfo {year}
  {2018})}\BibitemShut {NoStop}%
\bibitem [{\citenamefont {Esser}\ \emph {et~al.}(2016)\citenamefont {Esser},
  \citenamefont {Merolla}, \citenamefont {Arthur}, \citenamefont {Cassidy},
  \citenamefont {Appuswamy}, \citenamefont {Andreopoulos}, \citenamefont
  {Berg}, \citenamefont {McKinstry}, \citenamefont {Melano}, \citenamefont
  {Barch}, \citenamefont {di~Nolfo}, \citenamefont {Datta}, \citenamefont
  {Amir}, \citenamefont {Taba}, \citenamefont {Flickner},\ and\ \citenamefont
  {Modha}}]{esser:2016}%
  \BibitemOpen
  \bibfield  {author} {\bibinfo {author} {\bibfnamefont {S.~K.}\ \bibnamefont
  {Esser}}, \bibinfo {author} {\bibfnamefont {P.~A.}\ \bibnamefont {Merolla}},
  \bibinfo {author} {\bibfnamefont {J.~V.}\ \bibnamefont {Arthur}}, \bibinfo
  {author} {\bibfnamefont {A.~S.}\ \bibnamefont {Cassidy}}, \bibinfo {author}
  {\bibfnamefont {R.}~\bibnamefont {Appuswamy}}, \bibinfo {author}
  {\bibfnamefont {A.}~\bibnamefont {Andreopoulos}}, \bibinfo {author}
  {\bibfnamefont {D.~J.}\ \bibnamefont {Berg}}, \bibinfo {author}
  {\bibfnamefont {J.~L.}\ \bibnamefont {McKinstry}}, \bibinfo {author}
  {\bibfnamefont {T.}~\bibnamefont {Melano}}, \bibinfo {author} {\bibfnamefont
  {D.~R.}\ \bibnamefont {Barch}}, \bibinfo {author} {\bibfnamefont
  {C.}~\bibnamefont {di~Nolfo}}, \bibinfo {author} {\bibfnamefont
  {P.}~\bibnamefont {Datta}}, \bibinfo {author} {\bibfnamefont
  {A.}~\bibnamefont {Amir}}, \bibinfo {author} {\bibfnamefont {B.}~\bibnamefont
  {Taba}}, \bibinfo {author} {\bibfnamefont {M.~D.}\ \bibnamefont {Flickner}},\
  and\ \bibinfo {author} {\bibfnamefont {D.~S.}\ \bibnamefont {Modha}},\
  }\bibfield  {title} {\bibinfo {title} {Convolutional networks for fast,
  energy-efficient neuromorphic computing},\ }\href
  {https://doi.org/10.1073/pnas.1604850113} {\bibfield  {journal} {\bibinfo
  {journal} {Proceedings of the National Academy of Sciences}\ }\textbf
  {\bibinfo {volume} {113}},\ \bibinfo {pages} {11441} (\bibinfo {year}
  {2016})},\ \Eprint
  {https://arxiv.org/abs/https://www.pnas.org/content/113/41/11441.full.pdf}
  {https://www.pnas.org/content/113/41/11441.full.pdf} \BibitemShut {NoStop}%
\bibitem [{\citenamefont {Bellec}\ \emph {et~al.}(2018)\citenamefont {Bellec},
  \citenamefont {Salaj}, \citenamefont {Subramoney}, \citenamefont
  {Legenstein},\ and\ \citenamefont {Maass}}]{bellec:2018}%
  \BibitemOpen
  \bibfield  {author} {\bibinfo {author} {\bibfnamefont {G.}~\bibnamefont
  {Bellec}}, \bibinfo {author} {\bibfnamefont {D.}~\bibnamefont {Salaj}},
  \bibinfo {author} {\bibfnamefont {A.}~\bibnamefont {Subramoney}}, \bibinfo
  {author} {\bibfnamefont {R.}~\bibnamefont {Legenstein}},\ and\ \bibinfo
  {author} {\bibfnamefont {W.}~\bibnamefont {Maass}},\ }\bibfield  {title}
  {\bibinfo {title} {Long short-term memory and learning-to-learn in networks
  of spiking neurons},\ }in\ \href@noop {} {\emph {\bibinfo {booktitle}
  {Advances in Neural Information Processing Systems}}}\ (\bibinfo {year}
  {2018})\ pp.\ \bibinfo {pages} {787--797}\BibitemShut {NoStop}%
\bibitem [{\citenamefont {Neftci}\ \emph {et~al.}(2019)\citenamefont {Neftci},
  \citenamefont {Mostafa},\ and\ \citenamefont {Zenke}}]{neftci:2019}%
  \BibitemOpen
  \bibfield  {author} {\bibinfo {author} {\bibfnamefont {E.~O.}\ \bibnamefont
  {Neftci}}, \bibinfo {author} {\bibfnamefont {H.}~\bibnamefont {Mostafa}},\
  and\ \bibinfo {author} {\bibfnamefont {F.}~\bibnamefont {Zenke}},\ }\bibfield
   {title} {\bibinfo {title} {Surrogate gradient learning in spiking neural
  networks: Bringing the power of gradient-based optimization to spiking neural
  networks},\ }\href@noop {} {\bibfield  {journal} {\bibinfo  {journal} {IEEE
  Signal Processing Magazine}\ }\textbf {\bibinfo {volume} {36}},\ \bibinfo
  {pages} {51} (\bibinfo {year} {2019})}\BibitemShut {NoStop}%
\bibitem [{\citenamefont {Bradbury}\ \emph {et~al.}(2018)\citenamefont
  {Bradbury}, \citenamefont {Frostig}, \citenamefont {Hawkins}, \citenamefont
  {Johnson}, \citenamefont {Leary}, \citenamefont {Maclaurin},\ and\
  \citenamefont {Wanderman-Milne}}]{jax:2018}%
  \BibitemOpen
  \bibfield  {author} {\bibinfo {author} {\bibfnamefont {J.}~\bibnamefont
  {Bradbury}}, \bibinfo {author} {\bibfnamefont {R.}~\bibnamefont {Frostig}},
  \bibinfo {author} {\bibfnamefont {P.}~\bibnamefont {Hawkins}}, \bibinfo
  {author} {\bibfnamefont {M.~J.}\ \bibnamefont {Johnson}}, \bibinfo {author}
  {\bibfnamefont {C.}~\bibnamefont {Leary}}, \bibinfo {author} {\bibfnamefont
  {D.}~\bibnamefont {Maclaurin}},\ and\ \bibinfo {author} {\bibfnamefont
  {S.}~\bibnamefont {Wanderman-Milne}},\ }\href {http://github.com/google/jax}
  {\bibinfo {title} {{JAX}: composable transformations of {P}ython+{N}um{P}y
  programs}} (\bibinfo {year} {2018})\BibitemShut {NoStop}%
\bibitem [{\citenamefont {Wetterich}(2020)}]{wetterich2020probabilistic}%
  \BibitemOpen
  \bibfield  {author} {\bibinfo {author} {\bibfnamefont {C.}~\bibnamefont
  {Wetterich}},\ }\href@noop {} {\bibinfo {title} {The probabilistic world}}
  (\bibinfo {year} {2020}),\ \Eprint {https://arxiv.org/abs/2011.02867}
  {arXiv:2011.02867 [quant-ph]} \BibitemShut {NoStop}%
\bibitem [{\citenamefont
  {Wetterich}(2018{\natexlab{a}})}]{wetterich:2018information}%
  \BibitemOpen
  \bibfield  {author} {\bibinfo {author} {\bibfnamefont {C.}~\bibnamefont
  {Wetterich}},\ }\bibfield  {title} {\bibinfo {title} {Information transport
  in classical statistical systems},\ }\href@noop {} {\bibfield  {journal}
  {\bibinfo  {journal} {Nuclear Physics B}\ }\textbf {\bibinfo {volume}
  {927}},\ \bibinfo {pages} {35} (\bibinfo {year} {2018}{\natexlab{a}})},\
  \Eprint {https://arxiv.org/abs/1611.04820} {arXiv:1611.04820} \BibitemShut
  {NoStop}%
\bibitem [{\citenamefont
  {Wetterich}(2018{\natexlab{b}})}]{wetterich:2018quantum}%
  \BibitemOpen
  \bibfield  {author} {\bibinfo {author} {\bibfnamefont {C.}~\bibnamefont
  {Wetterich}},\ }\bibfield  {title} {\bibinfo {title} {Quantum formalism for
  classical statistics},\ }\href@noop {} {\bibfield  {journal} {\bibinfo
  {journal} {Annals of Physics}\ }\textbf {\bibinfo {volume} {393}},\ \bibinfo
  {pages} {1} (\bibinfo {year} {2018}{\natexlab{b}})},\ \Eprint
  {https://arxiv.org/abs/1706.01772} {arXiv:1706.01772} \BibitemShut {NoStop}%
\end{thebibliography}%
\end{document}